\documentclass{emulateapj}
\usepackage{natbib}
\usepackage{url}
\usepackage[caption=false]{subfig}

\begin{document}
\title{Polarized Light Imaging of the HD 142527 Transition Disk with the Gemini Planet Imager: Dust around the Close-in Companion\footnotemark[*]}
\footnotetext[*]{This paper includes data obtained at the Gemini-South telescope using the Gemini Planet Imager.}
\author{Timothy J. Rodigas\altaffilmark{1,4}, Katherine B. Follette\altaffilmark{2}, Alycia Weinberger\altaffilmark{1}, Laird Close\altaffilmark{2}, Dean C. Hines\altaffilmark{3}}

\altaffiltext{1}{Department of Terrestrial Magnetism, Carnegie Institute of Washington, 5241 Broad Branch Road, NW, Washington, DC 20015, USA; email: trodigas@carnegiescience.edu}
\altaffiltext{3}{Space Telescope Science Institute, Baltimore, MD 21218, USA}
\altaffiltext{2}{Steward Observatory, The University of Arizona, 933 N. Cherry Ave., Tucson, AZ 85721, USA}
\altaffiltext{4}{Carnegie Postdoctoral Fellow}
%\altaffiltext{4}{Department of Physics and Astronomy, University of Rochester, Rochester, NY 14627-0171, USA}
%\altaffiltext{5}{Rockhurst University, 1100 Rockhurst Rd, Kansas City, MO 64110, USA}
%\altaffiltext{6}{University of Toronto, 50 St George St., Toronto, ON M5S 1A1, Canada}
%\altaffiltext{7}{School of Earth and Space Exploration, Arizona State University, PO Box 871404, Tempe, AZ 85287-1404, USA}
%\altaffiltext{8}{Large Binocular Telescope Observatory, University of Arizona, Tucson, AZ 85721, USA}
%\altaffiltext{9}{University of Virginia, Department of Astronomy, 530 McCormick Road, Charlottesville, VA  22903, USA}

\newcommand{\about}{$\sim$~}
\newcommand{\mj}{M$_{J}$}
\newcommand{\degrees}{$^{\circ}$}
\newcommand{\arcseconds}{$^{\prime \prime}$}
\newcommand{\asec}{$\arcsec$}
\newcommand{\fasec}{$\farcs$}
\newcommand{\lprime}{$L^{\prime}$}
\newcommand{\ks}{$Ks$~}
\newcommand{\mjyasec}{mJy/arcsecond$^{2}$}
\newcommand{\microns}{$\mu$m}

%\shorttitle{something}
\shortauthors{Rodigas et al.}

\begin{abstract}
When giant planets form, they grow by accreting gas and dust. HD 142527 is a young star that offers a scaled-up view of this process. It has a broad, asymmetric ring of gas and dust beyond \about 100 AU and a wide inner gap. Within the gap, a low-mass stellar companion orbits the primary star at just \about 12 AU, and both the primary and secondary are accreting gas. In an attempt to directly detect the dusty counterpart to this accreted gas, we have observed HD 142527 with the Gemini Planet Imager in polarized light at $Y$ band (0.95-1.14 \microns). We clearly detect the companion in total intensity and show that its position and photometry are generally consistent with the expected values. We also detect a point-source in polarized light that may be spatially separated by \about a few AU from the location of the companion in total intensity. This suggests that dust is likely falling onto or orbiting the companion. Given the possible contribution of scattered light from this dust to previously reported photometry of the companion, the current mass limits should be viewed as upper limits only. If the dust near the companion is eventually confirmed to be spatially separated, this system would resemble a scaled-up version of the young planetary system inside the gap of the transition disk around LkCa 15.
\end{abstract}

\section{Introduction}
HD 142527 (spectral type F6 IIIe, \citealt{hd142527spectype}; age = 5$\pm$1.5 Myr, distance = 140$\pm$20 pc, \citealt{hd142527parameters}) is a young Herbig Ae/Be star with a complex circumstellar environment. It hosts a wide circumstellar disk, consisting of both dust and gas, located beyond \about 100 AU \citep{hd142527discovery}. Within 100 AU, the dust and gas density decline rapidly, revealing an apparent gap. An inner disk is also thought to exist beyond \about 5-10 AU, but its outer extent is not well-constrained \citep{hd142527midir,hd142527newsed}. The surface of the outer disk may contain water ice \citep{honda}--thought to be an essential ingredient for giant planet formation \citep{kokubo}. Recently \cite{hd142527biller} interferometrically detected a low-mass stellar companion within the inner disk, and \cite{hd142527close} directly imaged the \about 0.25 M$_{\odot}$ companion using Magellan adaptive optics (MagAO) by detecting a strong H$\alpha$ emission line that indicates gas accretion. Interestingly, \cite{hd142527streamers} detected gaseous streamers that appear to be crossing the disk gap on the opposite side of the companion, suggesting that both the primary and secondary stars are accreting gas from inside the gap, where warm gas is thought to exist \citep{hd142527warmgas}.

These recent findings raise several important questions: Is the secondary responsible for creating the wide gap? Since it is accreting gas, does it also have a \textit{circumsecondary} disk of dust? If it is surrounded by dust, did the dust originate in the inner or outer disk? To address these questions, high-contrast imaging capable of detecting dust at $<$ 0\fasec 1 is required. In this Letter, we report the direct detection of HD 142527B at $Y$ band (0.95-1.14 \microns) in total intensity, along with an offset source of polarized light, suggesting that dust is falling onto or orbiting the companion.  

% Specifically, to detect dust at such small inner working angles, precise point spread function (PSF) subtraction is required. Polarized differential imaging (PDI), in combination with adaptive optics (AO), has been shown to achieve high signal-to-noise (S/N) detections at very small inner working angles (e.g., \citealt{hd142527vltpol2,quanz169142,quanz100546}).

%We have obtained Gemini Planet Imager (GPI; \citealt{gpi}) polarized light images of HD 142527  We detect the low-mass stellar companion, HD 142527B, in total intensity at S/N \about 14. More importantly, we detect polarized light from the circumsecondary environment at S/N \about 11, suggesting 

%In Section \ref{sec:obs}, we describe the observations and data reduction. In Section \ref{sec:results}, we present our results. In Section \ref{sec:discussion}, we summarize our results and discuss their implications for the circumbinary environment of HD 142527. 

\section{Observations and Data Reduction}
\label{sec:obs}
\begin{figure*}[t]
\centering
\subfloat[]{\label{fig:PDI}\includegraphics[scale=0.45]{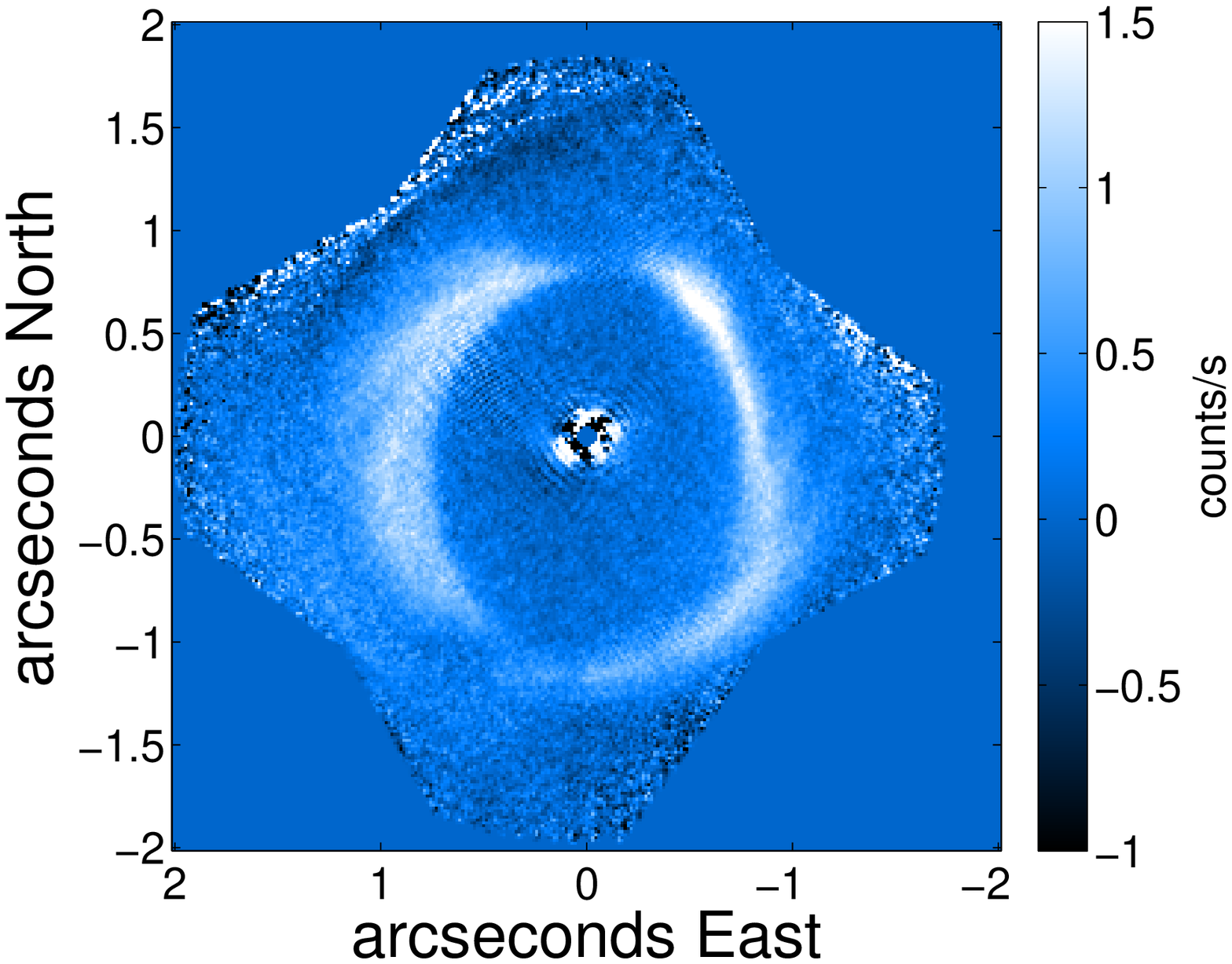}} 
\subfloat[]{\label{fig:PDIPAR}\includegraphics[scale=0.45]{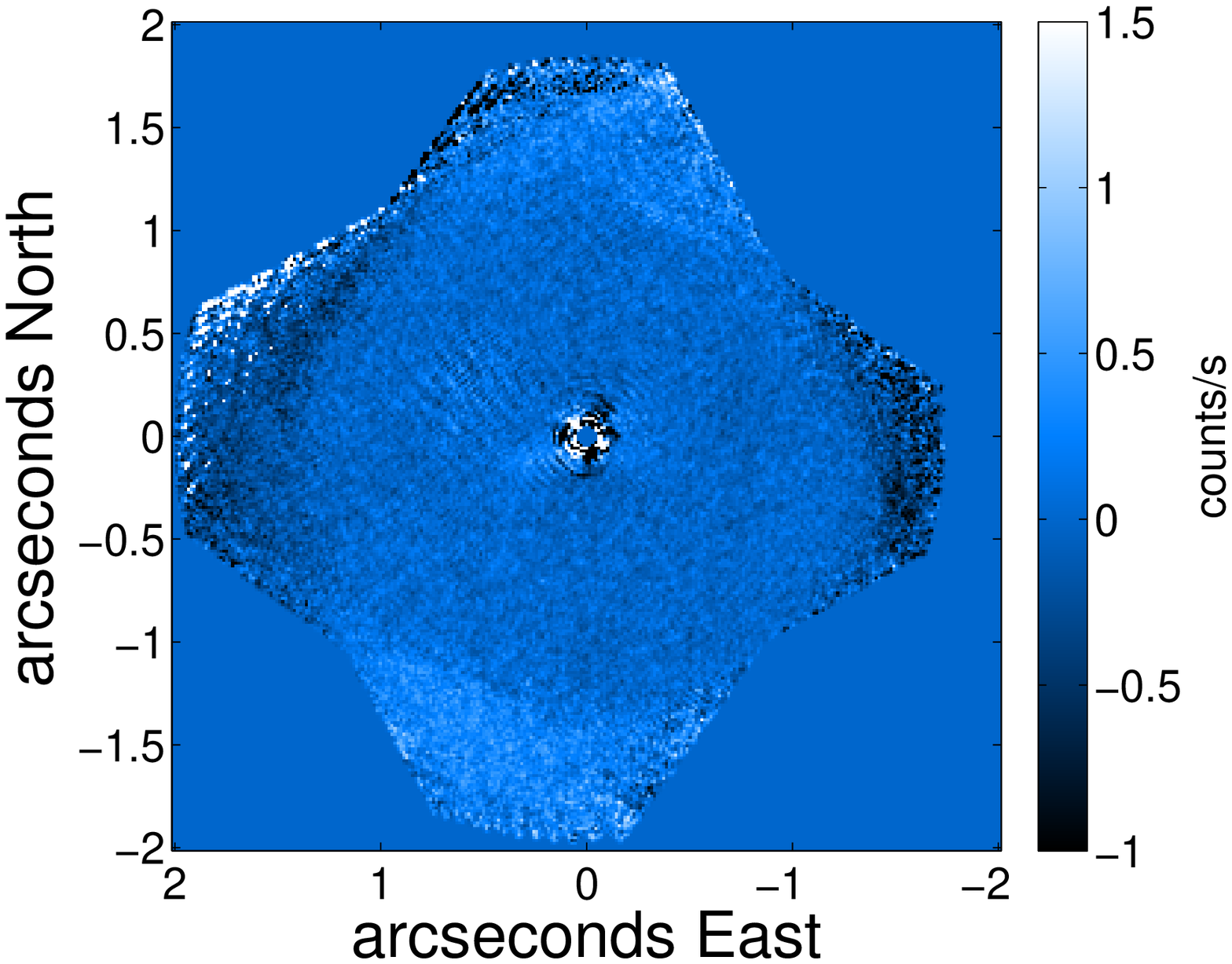}} \\
\subfloat[]{\label{fig:ADI}\includegraphics[scale=0.45]{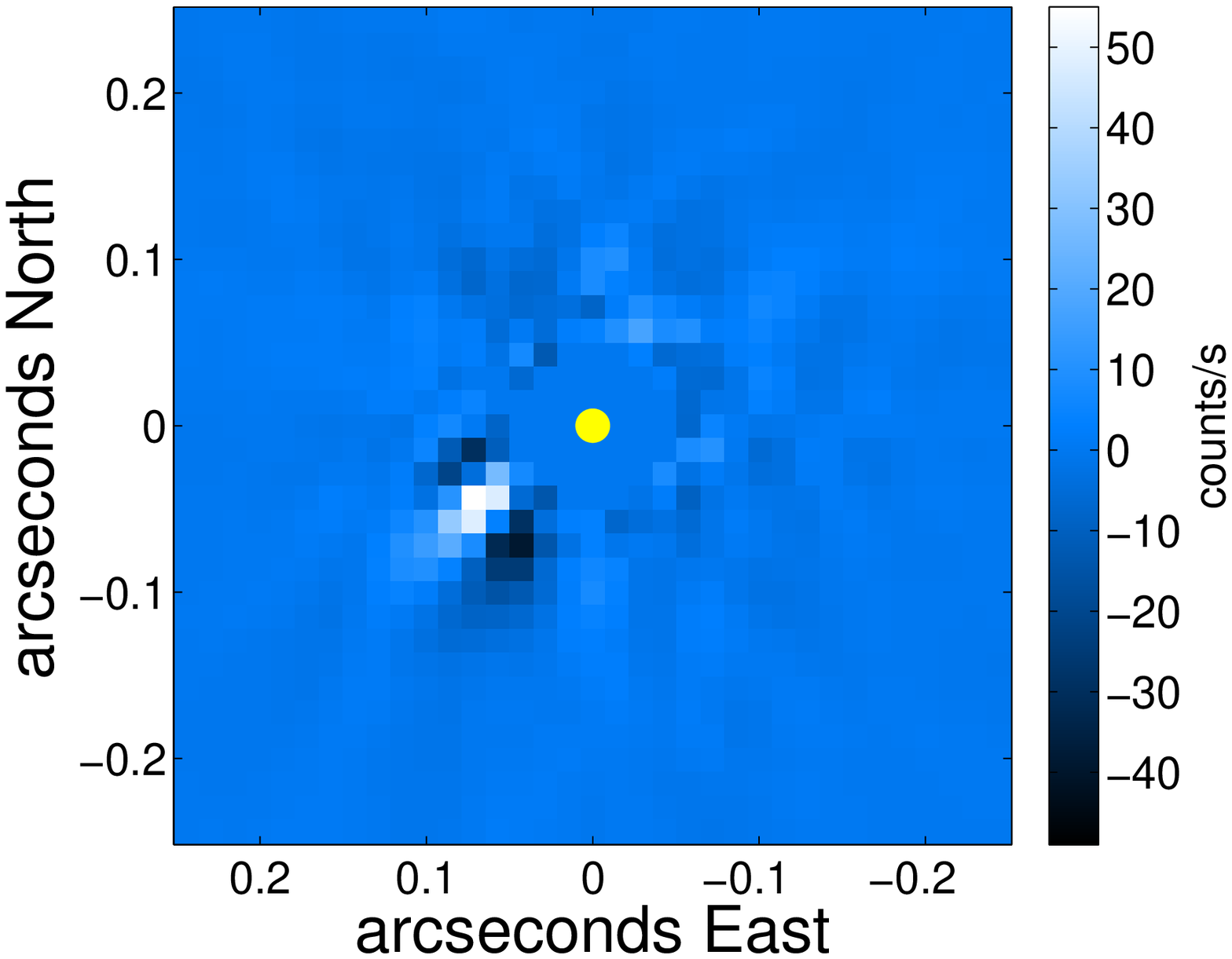}} 
\subfloat[]{\label{fig:ADISN}\includegraphics[scale=0.45]{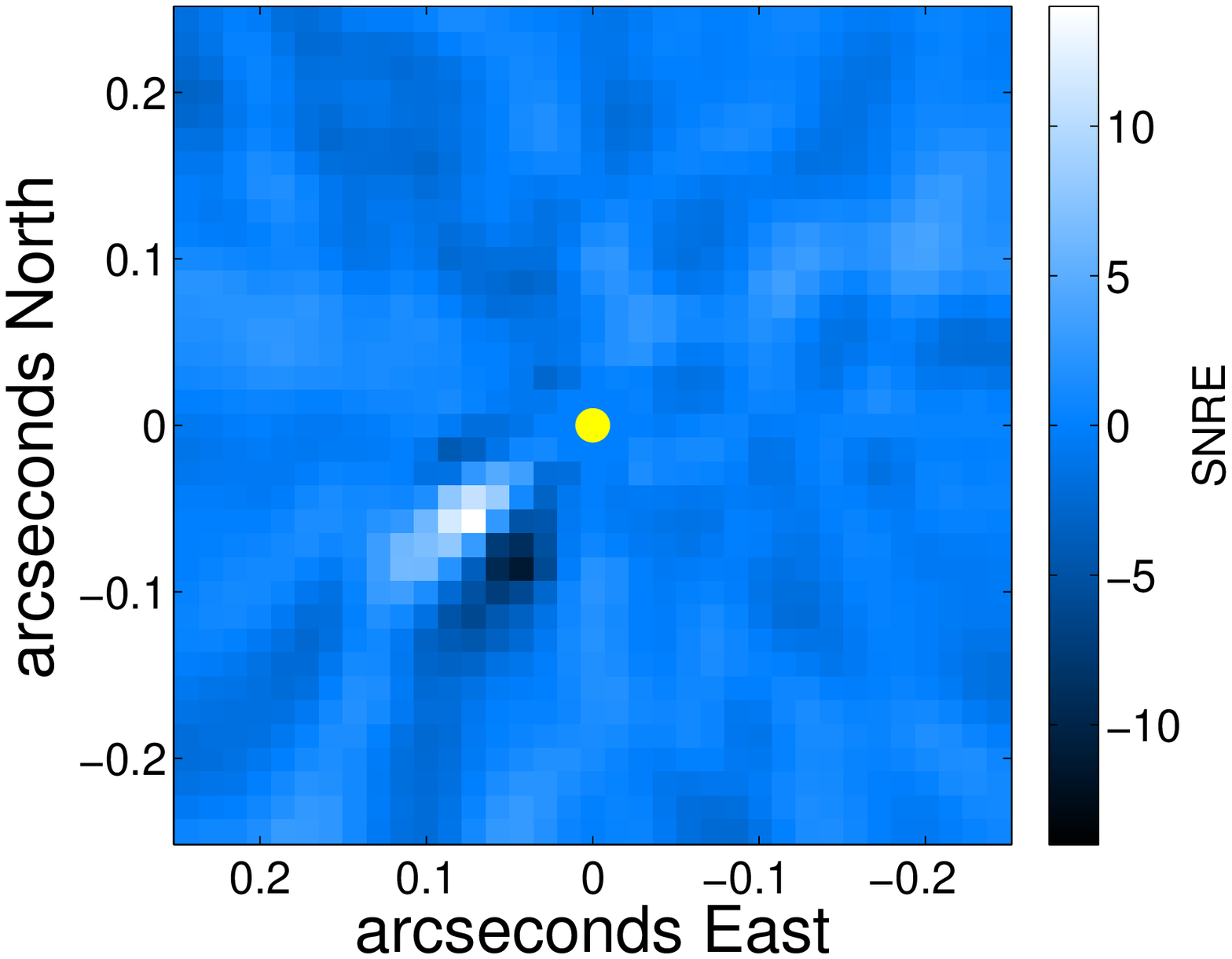}} \\
\caption{(a): PDI-processed $P_{\bot}$ image of the outer disk at $Y$ band. The outer disk is easily recovered. (b): PDI-processed $P_{\|}$ image, which contains \about zero signal at the location of the outer disk, as expected. (c): Zoomed-in total intensity image, obtained by reducing the data using ADI+PCA. The known companion, HD 142527B, is detected near its expected location. The negative residuals on either side of the companion are caused by self-subtraction in the ADI+PCA reduction. The yellow dot in this and other figures represents the location of the primary star. (d): SNRE map showing that the companion is detected at S/N \about 14.}
\label{fig:normal}
\end{figure*}

\begin{figure}[h!]
\centering
\subfloat[]{\label{fig:PDIzoom}\includegraphics[scale=0.42]{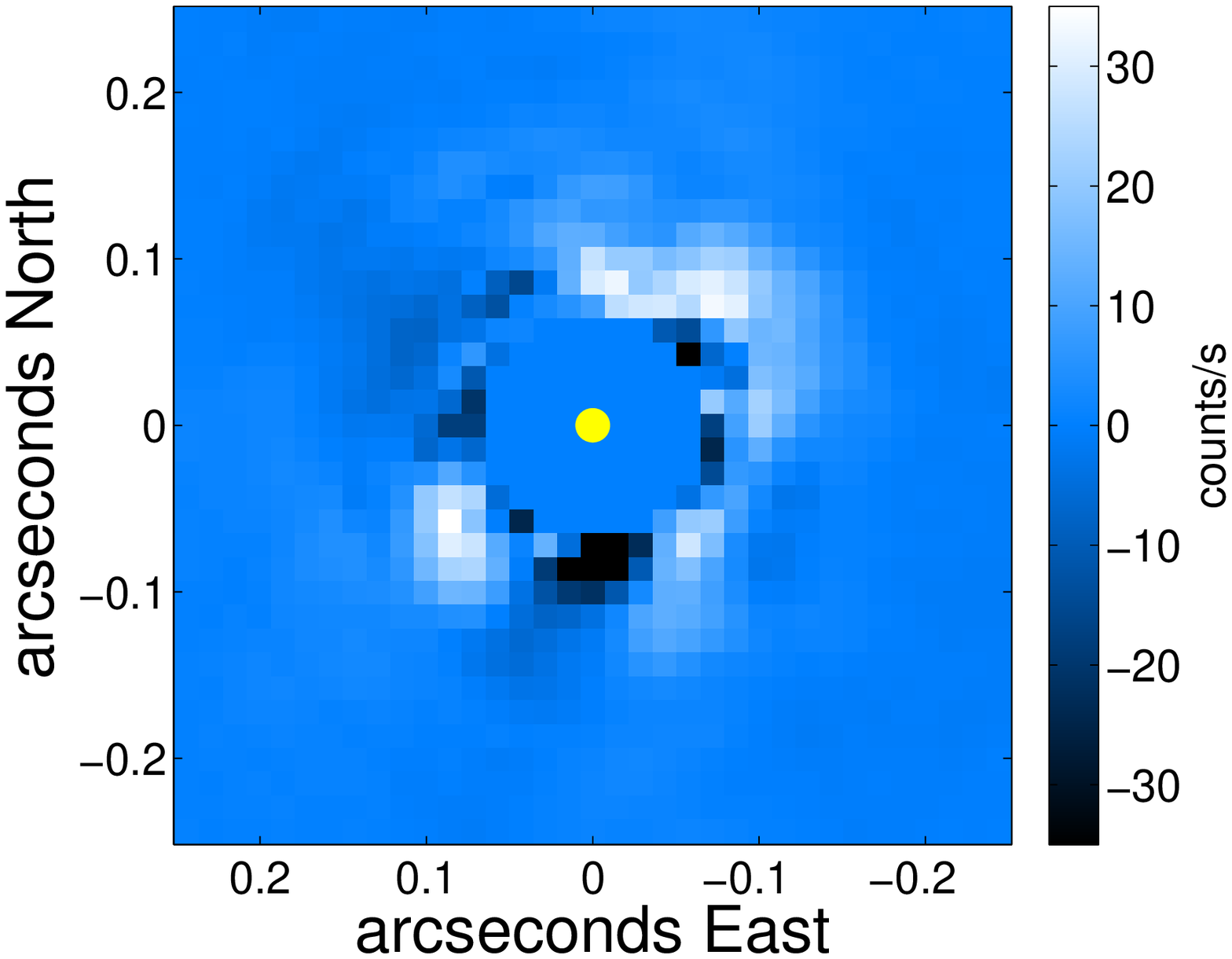}} \\
\subfloat[]{\label{fig:ADIPDI}\includegraphics[scale=0.42]{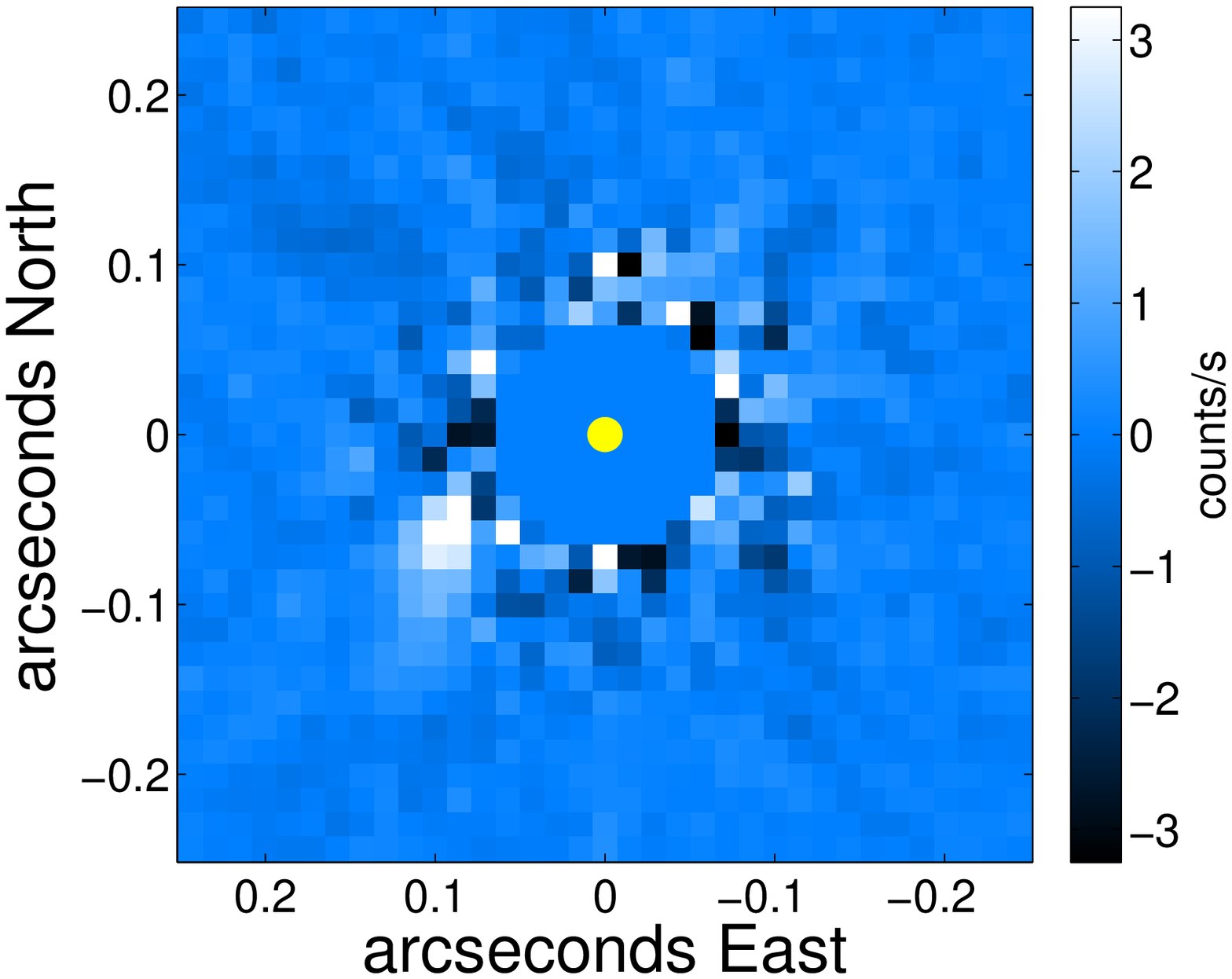}} \\
\subfloat[]{\label{fig:ADIPDISN}\includegraphics[scale=0.42]{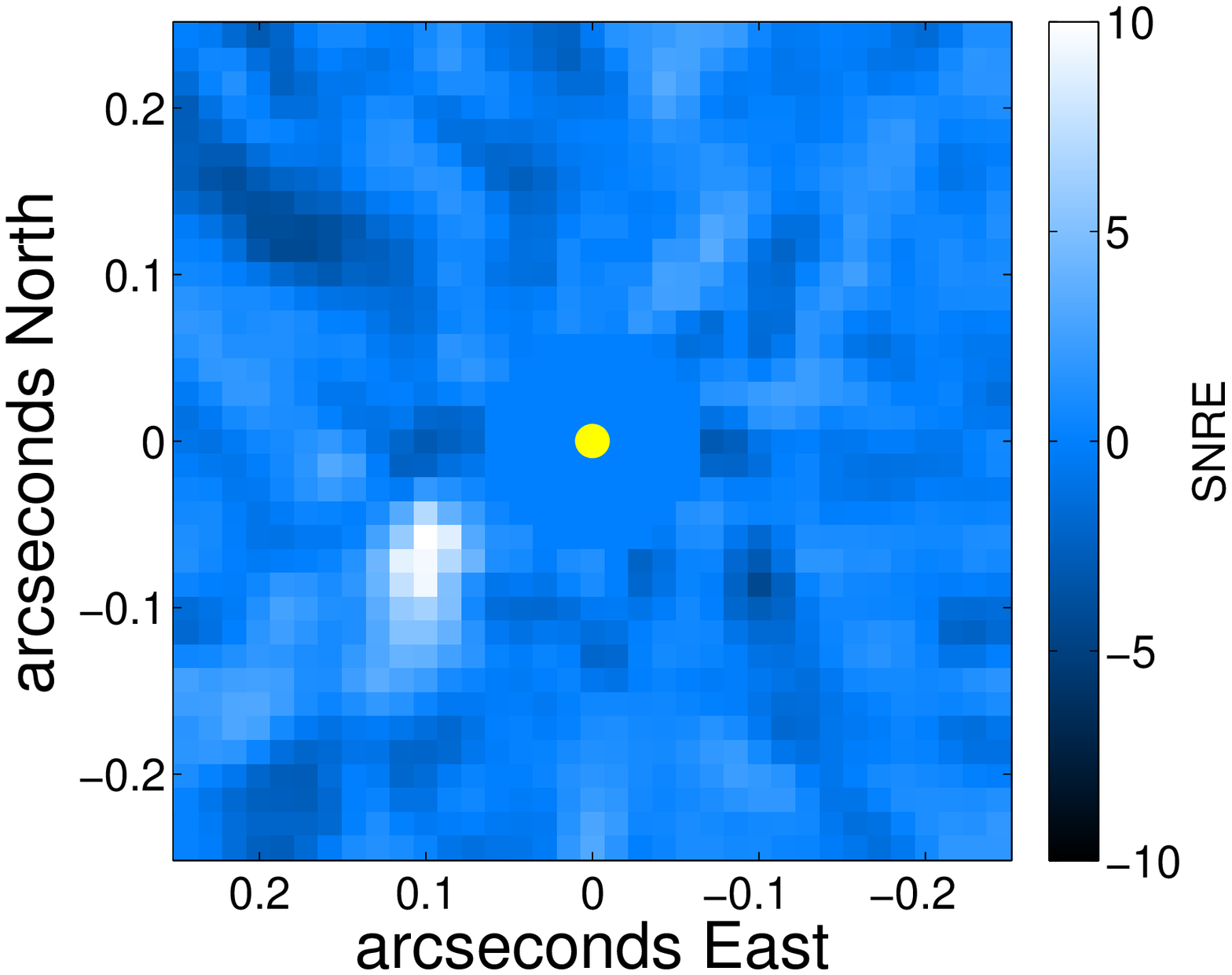}} \\
%\subfloat[]{\label{fig:ADIPDIPAR}\includegraphics[scale=0.42]{ADIPDIPAR.eps}} \\
\caption{(a): Final zoomed-in $P_{\bot}$ image generated using ODD, showing a bright point-source near the location of HD 142527B. (b): $P_{\bot}$ image, after additional PSF subtraction using ADI+PCA. The source is clearly detected and all other artifacts are removed. (c): SNRE map, showing that the polarized source is detected at S/N \about 11.}
\label{fig:ADIPDIimages}
\end{figure}

Observations of HD 142527 were carried out using the Gemini Planet Imager (GPI; \citealt{gpi}) on the Gemini-South Telescope at Cerro Pachon, Chile on the night of UT 25 April 2014 as part of the Early Science campaign. Images were obtained in polarized light at $Y$ band. GPI has a plate scale of 14.14 $\pm$ 0.01 mas/pixel \citep{gpiplatescale} and a field of view of 2\fasec 8 on a side. Observing conditions were good, with seeing at or below 0\fasec 75 for most of the night. To minimize the radial extent of saturation on the detector, we used the shortest possible integration time of 1.5 s and coadded 10 of these frames (15 s total per image). One image was obtained at each of the following half-wave plate angles: 0\degrees, 45\degrees, 22\degrees, 67\degrees\footnote{At the time of the observations, the Gemini Phase II software was incompatible with half-angle increments such that 22.5\degrees ~and 67.5\degrees ~were set to integer values.}, and then the sequence was repeated until 50 images were obtained, resulting in a total integration of 12.12 minutes. The instrument pupil was fixed during the observations, resulting in a total field of view rotation on the detector of 29.75\degrees ~and enabling angular differential imaging (ADI; \citealt{adi}). A PSF reference star, HIP 82885, was imaged immediately afterwards using the same instrumental setup for a total integration of 2.9 minutes. Images of HD 142527 were saturated within 0\fasec 05; images of the PSF reference star were not saturated (though some pixels were in the non-linear regime\footnote{http://www.gemini.edu/sciops/instruments/gpi/instrument-performance/detector-characteristics}). 

We used the GPI IDL data reduction package (v. 1.1)\footnote{http://docs.planetimager.org/pipeline/} to split each raw image into a datacube consisting of the ordinary and extraordinary beams. The processing pipeline locates the two polarization spots created by each lenslet and then constructs one image for each orthogonal polarization state. It also corrects for bad pixels and destripes the images.

Next we processed the images using our custom high-contrast imaging reduction routines written in Matlab. We divided each image by the number of coadds and the integration time to obtain units of counts/s. We then registered each image by calculating the center of light while ignoring saturated pixels. At this point, the dataset consisted of 50 ordinary beam images and 50 extraordinary beam images. 

To confirm the known outer disk structure, we followed the ``double ratio" polarized differential imaging (PDI) method (see \citealt{hd142527vltpol2} and references therein for more details), which yields $P_{\bot}$, the tangential polarization flux, and $P_{\|}$, the radial polarization flux. For this near face-on disk, $P_{\|}$ should contain \about zero signal and is thus a measure of the noise. After constructing these image sets, we rotated the images by their parallactic angles plus an instrumental offset to obtain North-up, East-left. To compute this offset, we reduced GPI spectral calibration data taken earlier in the April observing run on the Theta 1 Ori multiple star system at $H$ band. Using the known PA of the B$_{2}$-B$_{3}$ component from \cite{lairdtrapmagao}, which has very little proper motion over timescales of \about 1 year, we determined that an additional 3\degrees ~of clockwise rotation was needed to align the images with true North\footnote{\cite{gpiplatescale} found a rotational offset closer to \about 1\degrees; however, this was not computed using the Theta 1 Ori system.}. Fig. \ref{fig:PDI} and Fig. \ref{fig:PDIPAR} show the final $P_{\bot}$ and $P_{\|}$ images, respectively. The outer disk is clearly detected at high S/N, and no similar structures are evident in the radial polarization image. Furthermore, the known polarization ``holes" at position angles (PA) of \about 0\degrees ~and 160\degrees ~\citep{hd142527vltpol2,hd142527vltpol} are recovered at approximately the same locations.

\subsection{Recovering the companion in total intensity}
To detect HD 142527B in total intensity, we first added the 50 ordinary beam images with the 50 extraordinary beam images, yielding 50 total intensity images. We then reduced this dataset using our custom ADI pipeline in combination with Principal Component Analysis (PCA; \citealt{pca}). We varied the number of modes used to generate the PSFs, rotated the images by their parallactic angles to obtain North-up, East-left, and combined them using a mean with sigma clipping. The final number of modes was 8 (out of 50), since this resulted in a point-source being detected at a maximum S/N per resolution element (SNRE)\footnote{This was computed by convolving the final image with a Gaussian of FWHM = 38.2 mas, masking out the companion and computing the standard deviations in 1 pixel wide annuli as a function of radius, then dividing the Gaussian-smoothed image by these noise values.} \about 14 near the expected location of HD 142527B, based on the astrometry from \cite{hd142527biller} and \cite{hd142527close}. The final total intensity image and the corresponding SNRE map are shown in Fig. \ref{fig:ADI} and Fig. \ref{fig:ADISN}, respectively. These images, in particular the SNRE map, show that the object is radially extended, suggesting that additional signal resides \textit{outside} the point-source.

%The companion is located very close to its expected position based on the noted clockwise orbital motion \citep{hd142527close}. Given the very high S/N of the detection and the close agreement between our recovered position and the previously published astrometry, we consider the object in our images to be HD 142527B.

\subsection{Recovering polarized light near the companion}
To determine whether any polarized light is being scattered from the circumsecondary environment of HD 142527B, we combined the ``double difference" method (e.g., \citealt{doubledifference,hinkley4796}) with the ``double ratio" method. Specifically, we used PCA optimization to improve the PSF subtraction step in the former method and then used the latter method to correct for imperfect alignment of the half-wave plate \citep{hd142527vltpol2}. We refer to the combined method employed here as Optimized Double Differencing (ODD), which we describe below.

Typically, double differencing starts by generating $Q$ and $U$ images, which themselves are the differences of the ordinary and extraordinary beam images taken at various half-wave plate angles. The ordinary and extraordinary beam images are taken simultaneously, so the unpolarized star light is a good representation of the PSF. To improve the PSF subtraction in this step, we generated an optimal PSF from all of the available extraordinary images at a given half-wave plate angle using PCA. In other words, for the 0\degrees ~half-wave plate angle, a given $Q$ image was generated using Eq. \ref{eqn:eq1},
\begin{equation}
Q_{i} = O^{0\hbox{\degrees}}_{i} - PSF(E^{0\hbox{\degrees}}_{1:K}),
\label{eqn:eq1}
\end{equation}
where $i \in [1,12]$, with 12 being the number of images in a given half-wave plate sequence, $O$ ($E$) refers to the ordinary (extraordinary) beam of the polarized image, $PSF$ is the optimal PSF generated by PCA from the appropriate extraordinary images, and $K$ is the number of modes used to construct the PSF (here = 12). $-Q$, $U$, and $-U$ were generated in a similar manner using the 45\degrees, 22\degrees, and 67\degrees ~half-wave plate angle images, respectively.\footnote{Because off-axis sources rotate throughout observations, the polarization signal also rotates. This means that double differencing will result in a potentially biased polarized signal, depending on the speed and magnitude of the sky rotation between images. In our case, this bias is expected to be small and is outweighed by the gain in final S/N using ODD.}

The final $Q$ and $U$ images were computed in the normal manner (e.g., $Q - (-Q) = 2Q$). We then used the $Q$ and $U$ images to calculate $P_{\bot}$ and $P_{\|}$, following the double ratio method \citep{hd142527vltpol2}. At this point, if we rotate and combine the $P_{\bot}$ images, we detect a bright point-source near the location of HD 142527B in total intensity (Fig. \ref{fig:PDIzoom}). However, we can improve this detection by taking advantage of the ADI setup of the instrument and employing additional PSF subtraction on the (unrotated) $P_{\bot}$ images. We again used PCA, this time with 3 modes (out of 12).\footnote{We note that the PSFs used in this step were generated from the images themselves. Since they contain the (rotating) companion in polarized light, the flux of any recovered signal will have been attenuated.} After PSF subtracting, we rotated the images by their parallactic angles to obtain North-up, East-left and combined the images using a mean with sigma clipping. The final image is shown in Fig. \ref{fig:ADIPDI}, wherein the same point-source originally seen in Fig. \ref{fig:PDIzoom} is now recovered at S/N \about 11 (Fig. \ref{fig:ADIPDISN}), and all other artifacts have been removed. To check that the recovered source was not an instrumental polarization artifact, we repeated the above reduction on the $P_{\|}$ images. The final image did not show any similar structures near the polarized source, validating the detection.

\section{Results}
\label{sec:results}
\subsection{Astrometry}
\begin{figure}[t]
\centering
\includegraphics[scale=0.45]{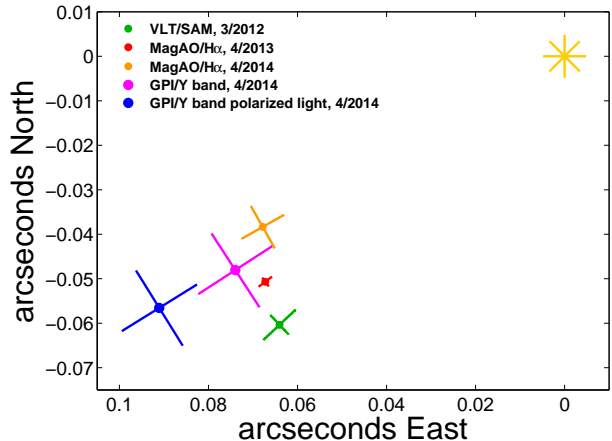}
\caption{Astrometry of the HD 142527 circumstellar environment. The blue and pink points are the polarized and total intensity locations, respectively, from this work. The green point is the location from \cite{hd142527biller} using Sparse Aperture Masking (SAM) with the VLT. The red point is the location from \cite{hd142527close} using MagAO imaging at H$\alpha$, and the orange point is the preliminary location from the same observations repeated in April 2014 (Follette et al., in prep.). The yellow asterisk marks the location of the primary star.} %The GPI location of HD 142527B in total intensity agrees with its expected location based on the previously observed clockwise orbital motion but is discrepant with the new MagAO location (by 1.4$\sigma$). The location of the polarized source is offset from the total intensity location by 2.7 AU (at $>$2$\sigma$ confidence).}
\label{fig:astrometry}
\end{figure}

To compute the astrometry of the point-source in total intensity (Fig. \ref{fig:ADI}) and polarized intensity (Fig. \ref{fig:ADIPDI}), we calculated the center of light in 3x3 pixel boxes centered on the brightest pixel near the location of the sources in each image. We assumed astrometric uncertainties of 0.5 pixels in the $x$ and $y$ directions, since GPI is just barely Nyquist-sampled at $Y$ band\footnote{$\lambda$/D at $Y$ band is 26.7 mas (1.9 pixels). By comparing the measured widths of inserted and recovered Gaussians with the width of the recovered companion, we found that the GPI PSF had a width of \about 38.2 mas (2.67 pixels). This satisfies the Nyquist requirement.}. In total intensity, the source has a separation from the primary of 88.25 $\pm$ 10.1 mas and PA of 123 $\pm$ 9.2\degrees. In polarized intensity, the source has a separation of 107.2 $\pm$ 10.1 mas and a PA of 121.84 $\pm$ 7.56\degrees. These locations are shown in Fig. \ref{fig:astrometry}, along with the previously published astrometry from 2012-2013 \citep{hd142527biller,hd142527close}. For reference we also plot the preliminary location of HD 142527B from MagAO/H$\alpha$ observations repeated in April 2014 (separation = 79.7$\pm$5.6 mas, PA = 119.5$\pm$ 8.17\degrees; Follette et al., in prep.). The location of the source in total intensity from GPI generally agrees with the expected location of HD 142527B based on orbital motion over the course of \about 1 year. It is marginally discrepant with the new MagAO location, though at 1.4$\sigma$ confidence. Thus we consider our source to be the companion. Interestingly, the projected separation of the polarized source is larger than that of the companion by \about 19.2 mas = 2.7 AU, at \about 2$\sigma$ confidence. However, this significance comes from the absolute astrometric uncertainty. Because both positions were measured in the same way using the same instrument, the \textit{relative} uncertainty is likely much smaller, meaning 2$\sigma$ is a lower limit.

\subsection{Companion photometry in total intensity}
%The ideal method for estimating the photometry of the companion in total intensity would be to insert and recover a scaled image of the (unsaturated) primary star at similar separations to the real companion, re-reduce the data, and iterate until a matching brightness was found. Unfortunately, no unsaturated images of the primary could be taken because HD 142527 is too bright ($V$ = 8.34), even in minimum exposure images. Fortunately, processed images of the PSF reference star, HIP 82885, were not saturated. However, the central \about few pixels were above 10,000 ADU, which is the non-linear threshold for GPI. The four unsaturated satellite spots could in principle be used, but their brightnesses relative to the central star have not been computed at $Y$ band, and their shapes are different enough from the central PSF that they are unsuitable for insertion and recovery. 

\begin{figure}[h]
\centering
\includegraphics[scale=0.45]{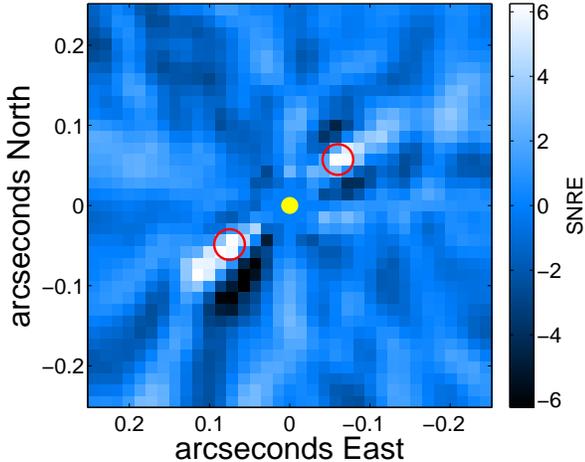}
\caption{SNRE map of the total intensity image including the inserted and recovered scaled PSF. The red circles mark the locations of the real and artificial sources. Residual signal is located outside the position of HD 142527B, whereas no similar signal is seen near the artificial source on the opposite side of the star.}
\label{fig:fakesn}
\end{figure}

We computed the companion photometry in total intensity by repeatedly inserting and recovering a scaled, mean-combined image of the unsaturated PSF, HIP 82885, at the same radius but \about 180\degrees ~away from the measured location of the companion. We varied the scale factor until the brightness of the artificial companion matched the measured brightness and S/N of the real companion (Fig. \ref{fig:fakesn}). Because the PSF image was non-linear in the central few pixels, we assumed a conservative uncertainty of 0.5 mag for the companion photometry. The optimal scale factor for the PSF was 0.055, corresponding to $\Delta Y_{PSF}$ \about 3.2. We used the stellar models from \cite{newkurucz} to fit the literature photometry of HIP 82885, including extinction given its distance of \about 850 pc \citep{van}. This yielded an apparent $Y$ mag of \about 8.2, which corresponds to the companion having $Y$ \about 11.4$\pm$ 0.5. This is marginally (1.7$\sigma$) fainter than the apparent magnitude of the companion at $H$ band (10.5$\pm$0.2; \citealt{hd142527biller}). 

We also note that Fig. \ref{fig:fakesn} shows that the companion is clearly extended in the radial direction, compared to the scaled artificial PSF inserted 180\degrees ~away. While not convincing on its own, this lends credence to the notion that an additional signal resides \textit{outside} the location of HD 142527B. 

\subsection{Polarization fraction near the companion}
We next computed the polarization fraction near the companion, since this can be informative of dust grain properties (\citealt{perrinabaur,grahampolarized}; and references therein). Typically one computes the ratio $p = P/I$, where $P$ is the image of the source in polarized light and $I$ is the image of the source in total intensity. Normally, for a disk of dust around a single central star, one performs PSF subtraction in total intensity to remove the star light and thus compute $I$ (e.g., \citealt{perrinabaur,hd142527vltpol2}). In our case, the polarized source is around the \textit{secondary}. This makes computing $p$ problematic because it is difficult to subtract the total intensity of the companion itself without also subtracting the circumsecondary source. Nonetheless, we can still compute a lower limit, $p_{min}$, recognizing that the companion's photospheric light is included. We accomplished this as follows: we considered $P$ to be our final polarized intensity image of the offset source (Fig. \ref{fig:ADIPDI}), and $I$ to be our final total intensity image of the companion (Fig. \ref{fig:ADI}). As Fig. \ref{fig:fakesn} shows, this image also contains scattered light extending away from the companion at the location of the polarized source. We computed $P/I$ and measured $p_{min}$ as the median in a 3x3 pixel box centered on the peak location of the polarized source. By inserting and recovering artificial sources, we verified that the attenuation due to ADI+PCA was comparable in both the $P$ and $I$ images, meaning no additional flux-correction factors were required. Therefore $p_{min} \approx 10\%$.

\section{Discussion and Summary}
\label{sec:discussion}
We have detected HD 142527B in total intensity at $Y$ band at approximately its expected location. Its brightness is marginally fainter than at $H$ band, perhaps suggesting that the companion is red at $Y-H$ like it is at 1.6-4 \microns ~\citep{hd142527biller}. We also detected a point-source in polarized light whose projected separation from the primary is larger than the companion's by \about 2.7 AU (at $>$2$\sigma$ confidence). The detection of polarized light is a strong indicator of scattering dust particles. This dust is mostly scattering light from the primary, since the $Y$ band flux received from the primary at the polarized source location is \about 3 times higher than the flux received from the companion. Assuming the polarized source is spatially separated from the companion, this system may resemble a scaled up version of the LkCa 15 system, which is thought to host a young protoplanet separated from one or more dust clumps \citep{krauslkca}.

The detection of polarized light near HD 142527B complicates its classification. Because the scattering dust is likely falling onto or orbiting the companion, it is possible that this dust contributes to the total intensity that has been directly detected at other wavelengths. If this is the case, the reported masses for the companion (0.1-0.4 M$_{\odot}$; \citealt{hd142527biller,hd142527close}) should be viewed as \textit{upper} limits only. To determine the lower limit on the companion mass, one would need to detect and subtract the contribution to the total light from the dust near the companion. Unfortunately, this is a daunting task because PSF subtraction on a faint \textit{secondary} at $<$ 0\fasec 1 is extremely difficult. 

Nonetheless, we can still make some inferences on the physical processes that would be occurring in the circumsecondary environment depending on the mass of the companion. Indeed the disk morphology and gap width depend uniquely on the binary mass ratio and orbital eccentricity \citep{eccbinarydisk}. If the companion is a low-mass star, then most of the light we see at $Y$ band is coming from the companion's photosphere, rather than the circumsecondary material. This in turn means that the true fractional polarization, $p$, would be much larger than $p_{min} \approx 10\%$. Since the outer disk also has large fractional polarizations (\about 20-50$\%$ \citep{hd142527vltpol2}), this could indicate that the dust near the companion is similar to the dust in the outer disk. Depending on the orbital eccentricity of the companion, which can be constrained in a few more years, the dust near the companion might have been swept up from the outer disk along its orbit. 

%For example, assuming the polarized light is coming solely from dust grains, we can determine the source of the light scattering off the dust. The primary has a luminosity of \about 20$L_{\odot}$ \citep{hd142527midir}, and the secondary has a luminosity of \about 0.01$L_{\odot}$, given a mass of \about 0.25$M_{\odot}$. Since the dust is \about 2 AU away from the companion and \about 12 AU away from the primary, the flux received from the primary is \about 55 times larger than the flux received from the secondary. This means most of the polarized light we are seeing originated from the primary star.

% that the companion is sculpting and clearing the outer disk, which is difficult to explain without invoking at least one other planetary mass companion farther out \citep{dodsondynamics,nelsondynamics}. 

If, on the other hand, the companion is a gas giant planet or protoplanet, then the light we see at $Y$ band (and at other wavelengths) is dominated by the circumsecondary dust, meaning the true fractional polarization would be small. This could indicate that the dust near the companion is \textit{different} from the dust in the outer disk. The dust near the companion might then originate from the inner disk. This would suggest that dust grains from the outer disk are not following the gap-crossing gas \citep{hd142527streamers}. It might also mean that multiple planets are responsible for the observed wide disk gap, since photoevaporation is unlikely in this accreting system \citep{dodsondynamics}.

%A third possibility that should be considered, given the large projected separation between the source of the polarized light and the companion, is that the companion is a low mass-star that is emitting bipolar jets, one of which we are seeing in polarized light. Jets are common features of young accreting stars, so it would not be unreasonable for this accreting object to have its own jets. In this case, the polarized light may originate from both dust and \textit{gas}, complicating the interpretations postulated above. 

This intriguing system and its multiple components must be monitored and observed at additional wavelengths (e.g., with GPI, MagAO, and/or SPHERE) so that these differing physical processes can be distinguished. The detection of the companion and nearby dust in polarized light in just 12 minutes of integration at $<$ 0\fasec 1 suggests that ODD in concert with ADI+PCA can be a powerful tool for detecting dust around off-axis point-sources. In particular, older planets with their own circumplanetary disks or rings may even be detectable by current facilities like GPI in the coming years.

%What is the source of the dust near the companion? ....say something about the dust properties from the polarization fraction limit....this might help us determine if the dust is coming from the inner or outer disk....

% accreting/dusty planets, planetary rings?
% origin of the dust

%\section{Summary}
%\label{sec:summary}

\acknowledgments
We are grateful to Marshall Perrin, Fredrik Rantakyro, Pascale Hibon, and the GPI science and instrument teams for their help obtaining, processing, and reducing the data. We thank the referee for helpful comments and suggestions. 

%\bibliographystyle{apj}
%\bibliography{ms}

\end{document}